\documentclass[12pt]{article}

\usepackage{bbm}
\usepackage{t1enc}
\usepackage[latin2]{inputenc}
\usepackage{amsmath,amssymb,amsthm}
\usepackage{txfonts}
\usepackage{sectsty}
\usepackage{overcite}
\usepackage{mathrsfs}

\def\Tr{\text{Tr }}

\def\l2{\mathbf{l}_2}
\def\unit{\mathbbm{1}}

\newcommand{\affiliation}[1]{
              \noindent\hspace*{2.5em}\parbox[t]{10cm}{\small\it#1}
              \vspace*{2ex}\\}

\theoremstyle{plain}\newtheorem{theor}{Theorem}
\theoremstyle{plain}\newtheorem*{prop}{Proposition}

\renewcommand{\leq}{\leqslant}
\renewcommand{\le}{\leqslant}
\renewcommand{\ge}{\geqslant}
\renewcommand{\geq}{\geqslant}

\renewcommand{\title}[1]{{\Large\bf\flushleft{#1}}\vspace*{3ex}\\}
\renewcommand{\author}[2]{{\noindent\hspace*{2.5em}\large#1}
                     \footnote{Electronic mail: $\mathtt{#2}$}\\}

\sectionfont{\normalsize\bf}
\subsectionfont{\normalsize\bf}
\makeatletter\renewcommand\@biblabel[1]{$^{#1}$}\makeatother

\begin{document}
\title{The von Neumann entropy asymptotics in multidimensional fermionic systems}

\author{S.~Farkas}{farkas@uchicago.edu}
\affiliation{Department of Physics, University of Chicago, Chicago,\\
             Illinois 60637}

\author{Z.~Zimbor\'as}{zimboras@rmki.kfki.hu} 
\affiliation{
Research Institute of Particle and Nuclear Physics,\\
H-1525 Budapest, P.O. Box 49, Hungary}
\affiliation{
Theoretische Physik, Universit\"at des Saarlandes,\\
             Saarbr\"ucken 66041, Germany}

\setcounter{footnote}{0}
\vspace*{-3ex}
\begin{abstract}
We study the von Neumann entropy asymptotics of pure translation-invariant
qua\-si-free states of $d$-dimensional fermionic systems. 
It is shown that the entropic area law is violated by all these states: 
apart from the 
trivial cases, the entropy of a cubic subsystem with edge 
length $L$ cannot grow slower than $L^{d-1}\ln L$. As for 
the upper bound of the entropy asymptotics, the 
zero-entropy-density property of these pure states is the only limit: 
it is proven that arbitrary fast sub-$L^d$ entropy growth is achievable.
\end{abstract}

\section{Introduction}

The structure of reduced density matrices belonging to subsystems 
of lattice models has been the focus of several recent studies 
\cite{VLRK,CC,PEDC,FHM,W,FZ,GK,BCS,LDRH,JK,KM,EZ,LS,HP,KMSW,P,
ELR,A,R,AFOV,LRI}. 
Especially, the von Neumann entropy of the reduced density matrices,
defined as $S(\rho) \coloneqq  - \Tr \rho \ln \rho $ 
, and its growth with the size of the subsystem have attracted 
much attention. In the context of condensed matter physics, a
non-saturating entropy asymptotics of the ground states 
of one-dimensional lattice models was found to be an indicator 
of quantum criticality \cite{VLRK}, moreover, this asymptotics was connected
by conformal field theoretical methods to the central charge of the theory
\cite{CC}. 
For multidimensional lattice systems the situation proved to be more 
complicated, namely, the von Neumann entropy asymptotics of 
ground states was found to depend on the statistics of the fields
defining the model. For free bosonic Hamiltonians with nearest
neighbor hopping terms the ground-state entropy  
was shown to grow with the area of the surface of the subsystem 
(regardless of the criticality of the system)\cite{PEDC},
while for free fermionic Hamiltonians with similar hopping terms
a logarithmic correction to the area-law was found \cite{W,GK,BCS,LDRH}.  
The statistics of the fields plays also an important role in
the interpretation of the von Neumann entropy as a measure of
entanglement in quantum information theory. 
Such an interpretation does not work for fermionic 
systems \cite{M,BCW}, but in the case of bosonic and
spin systems the von Neumann entropy of a 
subsystem is a natural measure of entanglement if the whole system is
in a pure state.\cite{BBPS}
Another motivation comes from data compression and DMRG theory:
it is believed that for translation-invariant states the 
von Neumann entropy of the reduced density matrix is related to the 
dimension of the "essential subspace" of the restricted state\cite{LS,HP}. 
Finally, let us mention two mathematical conjectures 
about the entropy asymptotics. A long-standing conjecture,
called the zero-entropy-conjecture (see e.g. Refs. [\citeonline{FHM,FZ}]),
states that the entropy density $\lim_{L \to \infty} S_L/L^d$ of pure translation-invariant
states of lattice spin and fermionic systems 
is zero. Where $S_L$ denotes the von Neuamnn entropy of the $d$-dimensional cubic subsystem with edge length $L$. 
Recently, it has also been conjectured
that the type of the von Neumann algebra 
obtained as the weak closure of the 
$\rm{C}^*$-algebra belonging to
the left (or right) half of a spin chain in a given state 
depends also on the von Neumann entropy asymptotics of the state \cite{KMSW}.

For the above (and also many other) reasons 
the von Neumann entropy asymptotics of 
several one- and multidimensional states has been studied. 
The most investigated states are pure translation-invariant 
quasi-free states of bosonic and fermionic systems.
As mentioned previously, the entropy asymptotics in these 
two cases are quite different.
In the bosonic case an entropic area law was shown to hold \cite{PEDC}, 
while for certain gauge-invariant fermionic quasi-free 
states the entropy asymptotics of
cubic subsystems was shown to be
$L^{d-1} \ln L$ (where $L$ is the edge length of the
cubic subsystem, and $d$ is the dimensionality of the system), but
it was also hinted (see e.g. Refs. [\citeonline{W,GK}])  
that the conditions made on the Fermi surface
might exclude somewhat exotic, but physically relevant states.
In this article we prove 
that any nontrivial (gauge-) and
translation-invariant quasi-free state gives at least an
$L^{d-1} \ln L$ entropy asymptotics, but the entropy asymptotics
can also be much faster than $L^{d-1} \ln L$. Actually, we
prove, that the zero-entropy-density conjecture is sharp
in the sense that for any function $F_L$ that has a 
sub-$L^d$ asymptotics, one can find a state which has a faster asymptotics
than $F_L$. 

\section{Translation-invariant quasi-free states}

In this section we shortly recall a few facts about
quasi-free states in order to be self-contained.
For much more detailed treatments see Refs. [\citeonline{BR,AF,F}], where also
the proofs of the statements mentioned in this section can be found.

The observable algebra $\mathcal{A}_d$ 
of a fermionic system on the $d$-dimensional cubic lattice 
$\mathbb{Z}^d$ is the 
CAR algebra corresponding to the Hilbert space $\ell^2(\mathbb{Z}^d)$, i.e., 
it is the $\rm{C}^*$-algebra generated by $\unit$ and 
$\{c_{\mathbf{k}} \; : \; \mathbf{k}=(k_1,k_2,\dots,k_d) 
\in \mathbb{Z}^d \}$, satisfying the canonical anticommutation relations:
\begin{eqnarray}
c_{\mathbf{k}}c_{\mathbf{k}'}+c_{\mathbf{k}'}c_{\mathbf{k}}&=&0 , 
\nonumber \\
c^{*}_{\mathbf{k}}c_{\mathbf{k}'}+
c_{\mathbf{k}'}c^{*}_{\mathbf{k}}&=&\delta_{\mathbf{k},\mathbf{k}'}\unit. 
\nonumber
\end{eqnarray}
The translation automorphisms $\alpha_{i}:\mathcal{A}_d \to \mathcal{A}_d$
corresponding to the $d$ different
canonical unit translations of $\mathbb{Z}^d$ are given by 
\begin{eqnarray}
\alpha_{1} (c_{\mathbf{k}})=c_{\mathbf{k}+(1,0,0, \dots 0 ,0)} \; ,\quad
\alpha_{2} (c_{\mathbf{k}})=c_{\mathbf{k}+(0,1,0, \dots 0 ,0)}\;, \quad 
\cdots \quad
\alpha_{d} (c_{\mathbf{k}})=c_{\mathbf{k}+(0,0,0, \dots 0 ,1)} \; .\nonumber
\end{eqnarray}
A state $\omega$ on $\mathcal{A}_d$ is called translation-invariant 
if $\omega \circ \alpha_{i}=\omega$ for every $\alpha_{i}$.

Let $Q$ be a positive bounded operator on  $\ell^2(\mathbb{Z}^d)$
for which $0 \le Q \le \unit$ holds.
The gauge-invariant quasi-free state $\omega_{Q}$  
belonging to this operator is defined by the rule:
\begin{equation}
\omega_Q(c_{\mathbf{k}_1}^{*} \dots c_{\mathbf{k}_n}^{*} 
c_{\mathbf{l}_m} \dots c_{\mathbf{l}_1} ) = 
\delta_{m,n} {\rm det} \left( \left[ 
Q_{\mathbf{k}_i,\mathbf{l}_j} 
\right]_{i,j=1}^{n} \right) , \nonumber
\end{equation}
where $Q_{\mathbf{k}, \mathbf{l}}$ denotes the matrix elements 
$\langle \psi_{\mathbf{k}}, Q  \psi_{\mathbf{l}}\rangle$. Here
$\psi_{\mathbf{k}}$ is the characteristic function of the lattice point 
$\mathbf{k}$ (i.e. 
$\psi_{\mathbf{k}}(\mathbf{k'})=\delta_{\mathbf{k},\mathbf{k'}}$). 
$Q$ is called the symbol of the quasi-free state $\omega_{Q}$.

Translation-invariant quasi-free states
are characterized by certain integrable functions
on the torus $\mathbb{T}^d=\times_{i=1}^d S^1$. 
Let us parametrize the $d$-dimensional torus $\mathbb{T}^d$
by $[-\pi,\pi)^d$. A gauge-invariant quasi-free state 
$\omega_{Q}$ is translation-invariant if 
and only if there exists an integrable function 
$q: [-\pi,\pi)^d \to [0,1) $ such that 
\begin{equation}
Q_{\mathbf{k},\mathbf{l}}=\frac{1}{(2 \pi)^d} 
\int_{-\pi}^{\pi}da_1 \dots \int_{-\pi}^{\pi}da_d\; 
q(a_1, \dots,a_d) 
{\rm e}^{-i [(l_1-k_1)a_1+ \dots +(l_d-k_d)a_d]} . \nonumber
\end{equation}
Furthermore, this state is pure if and only if
$q$ is (almost everywhere) equal to the characteristic 
function of a measurable set $\mathbb{M} \subset \mathbb{T}^d$.
Such a quasi-free pure state belonging to the measurable set $\mathbb{M}$
will be denoted by $\omega_{\mathbb{M}}$. 
$\mathbb{M}$ is called the "Fermi sea",
while the boundary of the interior points of 
$\mathbb{M}$ is called the "Fermi surface"
of the state $\omega_{\mathbb{M}}$.

\section{Entropy asymptotics of quasi-free states}

Let $\rho_L$ denote the density matrix obtained by
restricting the quasi-free state $\omega_{Q}$
to the subalgebra corresponding to the lattice points
$\{1,2,\dots , L \}^d \subset \mathbb{Z}^d$.
The von Neumann entropy of the restricted state,
$S_L \coloneqq - \Tr \rho_L \ln \rho_L$, can be expressed in terms
of Q as \cite{AF}:
\begin{equation*}
S_L=- \Tr \left( Q_{L}\ln Q_{L} +(\mathbbm{1}-Q_{L}) 
\ln (\mathbbm{1}-Q_{L})\right),  
\end{equation*} 
where $Q_L$ is the restriction of $Q$ to the $L^d \times L^d$ 
submatrix corresponding to the points 
$\{1,2,\dots , L \}^d \subset \mathbb{Z}^d$. 
The inequality $-x\ln x-(1-x)\ln(1-x) \geq\! x(1-x) $, 
which holds for $0\leq x\leq 1$, implies that
\begin{equation*}
S_L \geq \Tr Q_L(\unit-Q_L).
\end{equation*}

In the case of a pure (gauge- and) translation-invariant quasi-free state
$\omega_{\mathbb{M}}$ the above lower bound can rewritten,
as shown in Refs. [\citeonline{FHM},\citeonline{W}], in the form:
\begin{equation}
S_L \geq \frac{1}{(2 \pi)^d}
\int_{-\pi}^{\pi}da_1\dots\int_{-\pi}^{\pi}da_d\prod_{i=1}^dk_L(a_i)
\Lambda_\mathbb{M}(a_1,\dots a_d). \label{q}
\end{equation}
The definitions of $k_L$ and $\Lambda_{\mathbb{M}}$ are the following:  
\begin{equation*}
k_L(a)=\frac{\sin^2La/2}{\sin^2 a/2}, \quad {\rm and} \quad 
\Lambda_\mathbb{M}(\mathbf{a})=|\mathbb{M}\setminus \mathbb{M}+\mathbf{a}|,
\end{equation*}
where $|\cdot|$ denotes the Lebesgue measure,
and for any ${\bf a}=(a_1,a_2, \dots a_d)$ $\mathbb{R}^d$-vector
$\mathbb{M}+{\bf a}$ is the image of
$\mathbb{M}$ after a translation of the points of the torus 
$\mathbb{T}^d=\times_{i=1}^d S^1$
defined by rotating the first $S^1$ by $a_1$, 
the second $S^1$ factor by $a_2$, and so on.
Hence the vectors $\mathbf{a}$ and 
$\mathbf{a}+(2 \pi n_1, 2\pi n_2, \dots, 2 \pi n_d)$, where 
$n_i \in \mathbb{Z}$, act 
on the torus in the same way. 
The lower bound \eqref{q}, which was first developed 
by Fannes, Haegeman, and Mosonyi  [\citeonline{FHM}], 
will be the starting point of both of our theorems.
  
\subsection{Lower bound on the entropy asymptotics}

One can immediately observe that if the symbol of a pure 
quasi-free is $0$ or $\unit$
(i.e., if $|\mathbb{M}|$=0 or 
$|\mathbb{M}|=|\mathbb{T}^d|$ ), then 
$S_L=0$. We show in this section that all the other pure 
translation-invariant quasi-free states of
$d$-dimensional fermionic systems have at least an
$L^{d-1}\ln L$ entropy asymptotics.

\begin{theor}
Let $\omega_{\mathbb{M}}$ be a pure 
(gauge- and) translation-invariant quasi-free state
for which $0< |\mathbb{M}| <|\mathbb{T}^d|$. 
The entropy growth $S_L$ of  $\omega_{\mathbb{M}}$
is bounded from below by $c L^{d-1}\ln L$ 
for some $c > 0$ (which depends on $\mathbb{M}$).
\end{theor}

\begin{proof}
The proof is divided into four steps. 
First we investigate general properties of $\Lambda_\mathbb{M}$. 
Then putting everything together, we obtain a lower bound for 
$\Lambda_\mathbb{M}$ by the aid of which the proof can be easily 
completed in the last step.

\vspace{5 mm}
\noindent\emph{1. Continuity and subadditivity of $\Lambda_\mathbb{M}$}\\
\noindent The continuity of $\Lambda_\mathbb{M}$ can be proven
from Stone's theorem. According to this theorem
the representation of the translations in $L^2(\mathbb{T}^d)$ 
given by $(U_\mathbf{a}\psi)(\mathbf{x})=\psi(\mathbf{x}+\mathbf{a})$ 
is continuous in the strong topology, hence in the weak topology as well. 
Let $\chi_{\mathbb{M}}$ be the characteristic function of $\mathbb{M}$.
The difference 
$\Lambda_\mathbb{M}(\mathbf{b})-\Lambda_\mathbb{M}(\mathbf{a})$ 
can be written as
\[
\begin{gathered}
\Lambda_\mathbb{M}(\mathbf{b})-\Lambda_\mathbb{M}(\mathbf{a})=
\int_{\mathbb{T}^d}\chi_{\mathbb{M}}(\mathbf{x})
(1-\chi_{\mathbb{M}}(\mathbf{x}+\mathbf{b}))-
\int_{\mathbb{T}^d}
\chi_{\mathbb{M}}(\mathbf{x})(1-\chi_{\mathbb{M}}(\mathbf{x}+\mathbf{a}))=\\
\int_{\mathbb{T}^d}\chi_{\mathbb{M}}(\mathbf{x})
(\chi_{\mathbb{M}}(\mathbf{x}+\mathbf{a})
-\chi_{\mathbb{M}}(\mathbf{x}+\mathbf{b}))=
\langle\chi_{\mathbb{M}},(U_\mathbf{a}-U_\mathbf{b})\chi_{\mathbb{M}}\rangle.
\end{gathered}
\]
Weak continuity of $U_\mathbf{a}$ implies that this expression 
goes to zero as $\mathbf{a}$ goes to $\mathbf{b}$, thus 
$\Lambda_\mathbf{M}$ is continuous. 

Next, for any
two translations $\mathbf{a}$ and $\mathbf{b}$ 
the following holds:
\[
\mathbb{M}\setminus(\mathbb{M}+\mathbf{a}+\mathbf{b})
\subset\left[\mathbb{M}\setminus(\mathbb{M}+\mathbf{a})\right]
\cup\left[(\mathbb{M}+\mathbf{a})\setminus(\mathbb{M}+\mathbf{a}+\mathbf{b})
\right].
\]
By monotony and translational invariance of the Lebesgue measure we 
obtain the subadditivity property:  
\begin{equation*}
\Lambda_\mathbb{M}(\mathbf{a}+\mathbf{b})\leq
\Lambda_\mathbb{M}(\mathbf{a})+\Lambda_\mathbb{M}(\mathbf{b}).
\end{equation*}

\vspace{5 mm}
\noindent\emph{2. Irrelevant and relevant directions of $\Lambda_\mathbb{M}$ }\\
\noindent The subspace 
$\{ \kappa \mathbf{a} :  \kappa \in \mathbb{R} \}$
generated by a vector $\mathbf{a}\in \mathbb{R}^d$ is called
an {\it irrelevant direction} (with respect to $\mathbb{M}$)
if $\Lambda_\mathbb{M}(\kappa\mathbf{a})=0$ 
for all $\kappa\ \in \mathbb{R}$, otherwise it is called 
a {\it relevant direction}. Subadditivity of $\Lambda_\mathbb{M}$ 
implies that vectors generating irrelevant directions form a 
vector space: 
\begin{equation*}
\Lambda_\mathbb{M}( \kappa (\alpha \mathbf{a} +\beta \mathbf{b})) \le
\Lambda_\mathbb{M}(\kappa \alpha \mathbf{a})+ 
\Lambda_\mathbb{M}(\kappa \beta \mathbf{b})=0. 
\end{equation*}
However, if 
$\mathbf{a}$ and $\mathbf{b}$ generate
relevant directions, then  a linear combination of 
$\mathbf{a}$ and $\mathbf{b}$ 
can generate either a relevant or an irrelevant direction.

It is easy to show that there exists at least one relevant direction of
$\Lambda_\mathbb{M}$ if 
$\mathbb{M}$ is nontrivial (i.e., if 
$0< |\mathbb{M}| < |\mathbb{T}^d|$). 
If all directions were irrelevant, then 
by definition, $\mathbb{M}$ would remain invariant (up to a zero measure set)
under any translation. 
In this case one could define    
a translation-invariant measure $\mu$ on every Lebesgue measurable 
set $\mathbb{L}$ by the formula 
$\mu(\mathbb{L}) := |\mathbb{M} \cap \mathbb{L} |$. 
According to Haar's theorem any translation-invariant measure
on the torus is equal to the Lebesgue measure times a constant, 
i.e., $\mu(\mathbb{M})=k |\mathbb{M}|$. If $k=0$ then 
$|\mathbb{M}|=|\mathbb{M} \cap \mathbb{M}|=\mu(\mathbb{M})=0$, 
if $k > 0$, then 
$|\mathbb{T}^d| = \mu(\mathbb{T}^d)/k =
|\mathbb{M} \cap\mathbb{T}^d |/k=|\mathbb{M} \cap \mathbb{M}|/k =
\mu(\mathbb{M})/k= |\mathbb{M}| $.

Let $\mathbf{e}_i$ denote the $i$th standard unit vector
of $\mathbb{R}^d$, $\mathbf{e}_1=(1,0,0, \dots 0)$, 
$\mathbf{e}_2=(0,1,0, \dots 0)$, etc.
Vectors of the form $\kappa \mathbf{e}_i$ act on the torus
$\mathbb{T}^d=\mathop{\times}_{i=1}^dS^1$ by rotating only the $i$th 
$S^1$ factor and leaving the other $S^1$ factors invariant. 
We call the one-parameter subspaces of the form
 $\{\kappa\mathbf{e}_i : \kappa \in \mathbb{R} \}$ 
{\it principal directions}.
It follows from the previous discussion that all 
principal directions cannot be irrelevant 
(if $\mathbb{M}$ is nontrivial). 
By a permutation of the 
$S^1$ factors, we can achieve that the first $m$ ($m>0$) standard
unit vectors each generate a relevant direction, 
while the last $d-m$ generate irrelevant directions.  

\vspace{5 mm}
\noindent\emph{3. Lower bound for $\Lambda_\mathbb{M}$}\\

First we show that for every fixed relevant direction
there exists a linear lower bound.
Let $\mathbf{a}$ be a vector for which 
$\Lambda_\mathbb{M}(\mathbf{a})>0$.
Continuity of $\Lambda_\mathbb{M}$ implies that  
there is an $\epsilon>0$ and a $c>0$ such that 
$\Lambda_\mathbb{M}(\nu\mathbf{a})>c$ for any  
$1- \epsilon \le \nu \le 1$.
Let us denote by $ \lfloor x \rfloor$ the "lower integer part" of $x$ 
($x \ge \lfloor x \rfloor $). 
Now, $1-\epsilon \le \lfloor 1/\lambda \rfloor \lambda \le 1$ 
holds if $0 < \lambda \le \epsilon$. 
Using the subadditivity of $\Lambda_\mathbb{M}$, we obtain 
$c  < \Lambda_\mathbb{M}( \lfloor 1/\lambda \rfloor \lambda \mathbf{a}) \le 
\lfloor 1/\lambda \rfloor  \Lambda_\mathbb{M}( \lambda \mathbf{a}) \le 
 \Lambda_\mathbb{M}( \lambda {\mathbf{a}})/\lambda$. 
Summarizing, for any $\mathbf{a}$ that generates a
relevant direction, there exist a $c>0$ and
an $\epsilon >0$ so that 
\begin{equation*}
\Lambda_\mathbb{M}( \lambda \mathbf{a}) < c \lambda \; \; \; \;
 \text{for} \; \;
0< \lambda < \epsilon. 
\end{equation*}

However, this is not enough for an estimate of the integrand 
in \eqref{q}, which is our goal. 
Next we have to show that there exists
a sufficiently large set of relevant directions.
As we have mentioned in the previous part of the proof,
we can assume that the first $m$ standard basis vectors 
$\{\mathbf{e}_i\}_{i=1}^{m}$ each generate a relevant direction.
This does not mean that a linear combination of them
also generates a relevant direction,  
but we can circumvent this problem by finding an 
$m$-dimensional subregion in which the positive linear combinations
(positive cone) of vectors have this property.

If for every choice of signs $\{s_{i}\}_{i=1}^m$
a vector (not equal to zero or any of the 
first $m$ standard basis vectors $\mathbf{e}_i$) 
is picked from the sets 
$\{\sum_{i=1}^m a_i (s_i \mathbf{e}_i) \;| \; a_i \ge 0 \}$,
then these vectors will linearly generate the whole $\mathbb{R}^m$
vector space spanned by the standard basis vectors 
$\{\mathbf{e}_{i}\}_{i=1}^m$ .
Therefore there is a choice of signs
$\{s_{i}\}_{i=1}^m$ such that any vector in the compact set
$V\coloneqq\{ (s_1 a_1,s_2 a_2, \dots, s_m a_m, 0, \dots, 0)  
\;| \; a_i \ge 0,\;\sum_i a_i^2=1 \}$
generates a relevant direction, since if such a choice did not 
exist, then all the directions (including the ones generated
by the vectors $\{\mathbf{e}_i\}_{i=1}^m$ ) would be irrelevant, 
which contradicts our assumption. 

For any relevant direction we have a linear lower bound for 
$\Lambda_\mathbb{M}$ if the translation is sufficiently small. 
Unfortunately, the prefactor and the validity region of the linear lower bound depend on the 
direction,
so for a global lower bound of $\Lambda_{\mathbb{M}}$ 
we have to get rid of this direction dependence. For this purpose, 
first consider the following function defined on $V$:
\[
s(\mathbf{v})\coloneqq\sup\left\{\;c\;|\;\exists\epsilon>0\;\mbox{so that}\;
\Lambda_\mathbb{M}(\lambda\mathbf{v})\geq c\lambda\;\mbox{for any}\;
\lambda\leq\epsilon\right\}.
\]
We show that if $s_-=\inf_{\mathbf{v}\in V}s(\mathbf{v})=0$, then there would 
exist an irrelevant generator in $V$ in contradiction to its definition,
therefore $s_-$ is positive. Since $V$ is compact, if $s_-=0$ then there is a 
sequence $\mathbf{v}_n\!\in\! V$, which is convergent, and 
$\lim_{n\to\infty}s(\mathbf{v}_n)\!=\!0$. Let the limit of $\mathbf{v}_n$ 
be $\mathbf{v}$. By subadditivity of $\Lambda_\mathbb{M}$ and the definition of
the function $s$, for any positive integer $k$ there is an index $n_k$ so that
$\Lambda_\mathbb{M}(\lambda\mathbf{v}_{n_k})<\lambda/k$
for any $\lambda$. By continuity of $\Lambda_\mathbb{M}$,
\[
\Lambda_\mathbb{M}(\lambda\mathbf{v})=\lim_{k\to\infty}\Lambda_\mathbb{M}
(\lambda\mathbf{v}_{n_k})\leq\lim_{k\to\infty}\frac{\lambda}{k}=0.
\]

Let $0<\sigma<s_-$. It is important that $\sigma$ is strictly smaller than 
$s_-$. We define a function on $V$ (whose $\sigma$-dependence is suppressed 
because $\sigma$ is fixed from now on):
\[
\epsilon(\mathbf{v})\coloneqq\sup\left\{\;\epsilon\;|\;\Lambda_\mathbb{M}
(\lambda\mathbf{v})\geq\sigma\lambda\;\mbox{if}\;\lambda\leq\epsilon\right\}.
\]
We show that $\epsilon_-\coloneqq\inf_{\mathbf{v}\in V}\epsilon(\mathbf{v})>0$. 
The argument is similar to the one we have just finished. Suppose the contrary.
$V$ is compact, so we have a convergent sequence $\mathbf{v}_n$, with limit 
$\mathbf{v}$, such that $\lim_{n\to\infty}\epsilon(\mathbf{v}_n)=0$. Note that our choice $\sigma<s_-$ guarantees that $\epsilon$ is strictly positive on $V$. Continuity of $\Lambda_\mathbb{M}$ implies that $\Lambda_\mathbb{M}(\epsilon(\mathbf{u})\mathbf{u})=\sigma\epsilon(\mathbf{u})$ for any $\mathbf{u}\in V$. Consequently, 
\[
\lim_{n\to\infty}\Lambda_\mathbb{M}\left(
\left\lfloor\frac{\lambda}{\epsilon(\mathbf{v}_n)}\right\rfloor\epsilon(\mathbf{v}_n)\mathbf{v}_n\right)\leq\lim_{n\to\infty}\left\lfloor\frac{\lambda}{\epsilon(\mathbf{v}_n)}\right\rfloor
\Lambda_\mathbb{M}\left(\epsilon(\mathbf{v}_n)\mathbf{v}_n\right)=\lim_{n\to\infty}
\left\lfloor\frac{\lambda}{\epsilon(\mathbf{v}_n)}\right\rfloor\epsilon(\mathbf{v}_n)\sigma=\sigma\lambda
\]
for any $\lambda$. But $\lim_{n\to\infty}\lfloor\lambda/\epsilon(\mathbf{v}_n)\rfloor\epsilon(\mathbf{v}_n)\mathbf{v}_n=\lambda\mathbf{v}$,
and $\Lambda_\mathbb{M}(\lambda\mathbf{v})>\sigma\lambda$ for some $\lambda$ 
(the latter inequality is strict, this is the point where our choice 
$\sigma<s_-$ comes into play again), which contradicts the continuity of 
$\Lambda_\mathbb{M}$.

At last we arrived at the advertised lower bound for $\Lambda_\mathbb{M}$:
\begin{equation}
\Lambda_\mathbb{M}(\mathbf{v})\geq
\sigma\parallel\!\mathbf{v}\!\parallel\;\;\;\;\mbox{if}\;\;\frac{\mathbf{v}}{\parallel\!\mathbf{v}\!\parallel}\in V,\;\;\mbox{and}\;\;\parallel\!\mathbf{v}\!\parallel<\epsilon_-.
\label{finalbound}
\end{equation} 

\vspace{5 mm}
\noindent\emph{4. A lower bound for the entropy asymptotics}\\
\noindent We can write the lower bound \eqref{q} as
\[
\begin{aligned}
S_L&\ge \frac{1}{(2 \pi)^d}\int_{-\pi}^{\pi}da_1\dots\int_{-\pi}^{\pi}
da_m\prod_{i=1}^mk_L(a_i)\Lambda_\mathbb{M}(P_R\mathbf{a})
\left(\int_{-\pi}^{\pi}da_{m+1}
\dots\int_{-\pi}^{\pi}da_d\prod_{i=d-m}^dk_L(a_i)\right)\\&\geq
\frac{L^{d-m}}{(2\pi)^m}
\left|\int_0^{s_1\epsilon_-/\sqrt{m}}da_1\dots\int_0^{s_m\epsilon_-/\sqrt{m}}da_m\;
\sigma\parallel\!P_R\mathbf{a}\!\parallel\right|\\
&\geq\frac{L^{d-m}}{(2 \pi)^m}\sigma\int_0^{\epsilon_-/\sqrt{m}}da_1\dots
\int_0^{\epsilon_-/\sqrt{m}}da_m\;a_1\prod_{i=1}^mk_L(a_i).
\end{aligned}
\] 

In the first inequality we simply used the fact that the irrelevant 
translations alter $\mathbb{M}$ only by a zero measure set, 
so in the argument of $\Lambda_\mathbb{M}$ the last $d-m$ components can be set
to zero 
($P_{R}$ is the standard projection from 
$\mathbb{R}^d=\mathbb{R}^m \times \mathbb{R}^{d-m}$ to the subspace 
$\mathbb{R}^m$ 
generated by the first $m$ standard unit vectors), 
and the integrations over the irrelevant principal directions can be 
pulled out. Next, these integrals were performed, and the integration region 
was shrunk into a hypercube where the lower bound \eqref{finalbound} can be 
applied. Finally, we replaced the Euclidean norm of 
$P_{R}\mathbf{a}$ with its first component. 

Borrowing the inequalities 
\[
\int_0^\delta da\;k_L(a)\geq c_1L,\;\;\;\;\int_0^\delta da\;a\;k_L(a)\geq c_2\ln L
\]
from Ref. \citeonline{FHM}, which are valid in the case $L>1$ for
some $c_1 >0$ and $c_2>0$,
the proof is complete:
\[
S_L\geq c L^{d-1}\ln L
\]
with some constant $c>0$.
\end{proof}

\subsection{No sub-$L^d$ upper bound  for the entropy asymptotics}
The result of the previous section was quite general: it 
holds for any translational-invariant pure quasi-free state. 
In Ref. [\citeonline{GK}] it was shown that 
for Fermi surfaces satisfying certain
conditions the entropy asymptotics is in fact $c' L \ln L$.
However, we show that the area law can be violated to a higher 
degree than logarithmic for general quasi-free states.
In a sense, it can be broken to any extent 
permitted by the zero-entropy-density conjecture (which is 
in fact a theorem for the quasi-free states). The crux of 
the proof of this statement is the following observation:
\begin{prop}
Let $h$ be a strictly monotonically increasing continuous function defined 
on the  interval $[0,\epsilon]$ ($\epsilon>0$). Furthermore, let $h(0)=0$. 
There exists a set $M\subset S^1$ ($d=1$) such that
\begin{equation}
\Lambda_M(a)\geq h(a) 
\label{h}
\end{equation}
for sufficiently small $a$. 
\end{prop}
\begin{proof}
See Ref. [\citeonline{FZ}].
\end{proof}
Actually, the states constructed by the aid of $M$ are 
not even so exotic; their Fermi sea is the union of (countably 
many) disjoint intervals. Now, we can repeat the proof of the theorem in 
Ref. [\citeonline{FZ}] almost verbatim to show that it can 
be generalized to any spatial dimension $d$.

\begin{theor}
Let $F:\mathbb{N}\to\mathbb{R}^+$ be a function that satisfies 
$\lim\nolimits_{L\to\infty}F_L/L^d=0$. There exists a pure quasi-free 
state such that $S_L\geq F_L$ for sufficiently large $L$.  
\end{theor}
\begin{proof}
Let us define the function $f_L \coloneqq F_L/L^d$. This statisfies
$\lim\nolimits_{L\to\infty}f_L/L=0$.
In Ref. [\citeonline{FZ}] we argued that 
$h(x)\coloneqq\frac{d}{dx}(xg(x))$ satisfies the 
conditions of the previous proposition if 
$g$ is a suitably chosen positive function for which
\[
\frac{2}{\pi^2}g\left(\frac{\pi}{L}\right)\geq \frac{f_L}{L}.
\]
Therefore there exists a set $M \ [-\pi, \pi)$ for which \eqref{h} 
holds with this particular $h$. 
Let $\mathbb{M}=\times_{i=1}^{d-1}[-\pi,\pi)\times M$. 
Then \eqref{q} simplifies to  
\begin{equation*}
S_L\geq \frac{1}{(2 \pi)^d}\left(\int_{-\pi}^{\pi}da\;k_L(a)\right)^{d-1}
\int_{-\pi}^{\pi}db\;k_L(b)\;\Lambda^{(1)}_M(b)=
\frac{L^{d-1}}{2 \pi}\int_{-\pi}^{\pi}db\;k_L(b)\; \Lambda^{(1)}_M(b).
\label{qsimple}
\end{equation*}
Restricting the integration region and 
using \eqref{h}, we obtain for sufficiently large $L$
the final inequality:
\[
S_L\geq 
\frac{L^{d-1}}{2 \pi}\int_{0}^{\pi/L}db\;k_L(b)\; h(b)
\geq \frac{L^{d-1}}{2 \pi}\int_{0}^{\pi/L}db\; \frac{4L^2}{\pi^2}\; h(b)
\geq \frac{2L^{d}}{\pi^2}g\left(\frac{\pi}{L}\right)=F_L
\]
\end{proof}
\section*{Acknowledgments}
We would like to thank P. Vecsernyés and K. Szlachányi for their useful 
suggestions. One of the authors (Z. Z.) was partially supported by the
German-Hungarian exchange program (DAAD-MÖB).

\end{document}